\documentclass[aps,pre,twocolumn,showpacs,superscriptaddress,groupedaddress]{revtex4-1}
\usepackage{graphicx}
\usepackage{bm}
\usepackage{dcolumn}
\usepackage{rotating}
\usepackage[colorlinks=true,linkcolor=blue,filecolor=blue,menucolor=yellow,urlcolor=blue,
citecolor=blue,anchorcolor=blue]{hyperref}
\usepackage{color}
\usepackage{subfig}
\usepackage{amsmath}
\usepackage{chemmacros}
\usepackage{filecontents}
\usepackage[titletoc,toc,title]{appendix}
\begin{document}


\title{Interstitial flows regulate collective cell migration heterogeneity through adhesion }

\author{Himadri S Samanta}\affiliation{Department of Chemistry, University of Texas at Austin, TX 78712}
\date{\today}
\begin{abstract}
The migration behaviors of cancer cells are known to be heterogeneous. However, the interplay between the adhesion interactions, dynamical shape changes and fluid flow in regulating cell migration heterogeneity and plasticity during cancer metastasis is still elusive. To further quantitative understanding of cell motility and morphology, we develop a theory using a stochastic quantization method that describes the role of biophysical cues in regulating diverse cell motility.
We show that the cumulative effect of time-dependent adhesion interactions that determine the structural rearrangements and self-generated force due to actin remodeling dictate the super-diffusive motion of mesenchymal phenotype in the absence of flow. Interstitial flows regulate cell motility phenotype and promote the amoeboid over mesenchymal motility through adhesion interactions. Cells exhibit a dynamical slowing down of collective migration, with a decreasing degree of super-diffusion. Mesenchymal cells are more persistent and diffusive compared to amoeboid cells. 
Our findings suggest a mechanism of Interstitial flow-induced directed motion of cancer cells through adhesion and provide the much-needed insight into a recent experimental observation concerning the diverse motility of breast cancer cells.
\end{abstract}

\maketitle
\def\s{\rule{0in}{0.32in}}

\section{Introduction}
Collective cancer cell invasion, followed by local and distant metastasis, is a hallmark of cancer\cite{Hanahan11Cell}.  Cancer metastasis is a multistep process, where tumor cells detach from primary tumor, invade through the interstitial extracellular matrix, intravasation of tumor cells into vascular vessels, extravasation of circulating tumor cells to peripheral tissues, and establish a secondary tumor at distant organ\cite{Sethi11NRC,Mitchell13FO,Chambers02NRC,Chaffer11Science,Friedl09NRMCB,Friedl04IJDB,Lecaudey06COCB,Roth07CB,Ilina09JCS,Friedl12NCB}. 
Dynamics associated with invasion and metastasis, involve the collective cell migration regulated by biomechanical (e.g. cytokines secreted by cells and nutrients) and biophysical cues (e.g. fluid flows and ECM)\cite{Hompland12CR,Heldin04NRC,Less92CR,Boucher96CR,Weidner91NEJM}.
Tumor cells reside in an extracellular matrix (ECM) containing interstitial fluid that transports nutrients and signaling molecules. The interstitial flow has been shown to affect the morphology and migration of cells such as fibroblasts, cancer cells, endothelial cells, and mesenchymal stem cells\cite{Swartz07ARBE}. Interstitial flows are particularly important for tumor cell invasion because it is elevated in tumor microenvironment due to the heightened interstitial fluid pressure as well as the abnormal angiogenic, lymphangiogenic blood and lymphatic vessels\cite{Munson13CR,Heldin04NRC,Less92CR,Boucher96CR,Weidner91NEJM,Swartz01MRT,Skobe01NM,Zaidel-Bar3997}. The flow speed associated with interstitial flows are in the order of a few micrometers per second in normal tissue\cite{Chary89PNAS,Hompland12CR,Munson14CMR,SHIELDS2007526}.
Two types of motility (e.g. amoeboid and mesenchymal) have been broadly categorized in cell migration in 3D architechture\cite{Friedl10JCB,Condeelis03NRC}. Cells with aspect ratio smaller than 2.0 are considered rounded or amoeboid and cells with an aspect ratio greater than 2.0 are considered elongated or mesenchymal (see fig.1)\cite{Wolf03JCB,Petrie12JCB, SANZMORENO08Cell}.
\begin{figure}[b]
\vspace{-.2 in}
	\includegraphics[width=.5\textwidth]{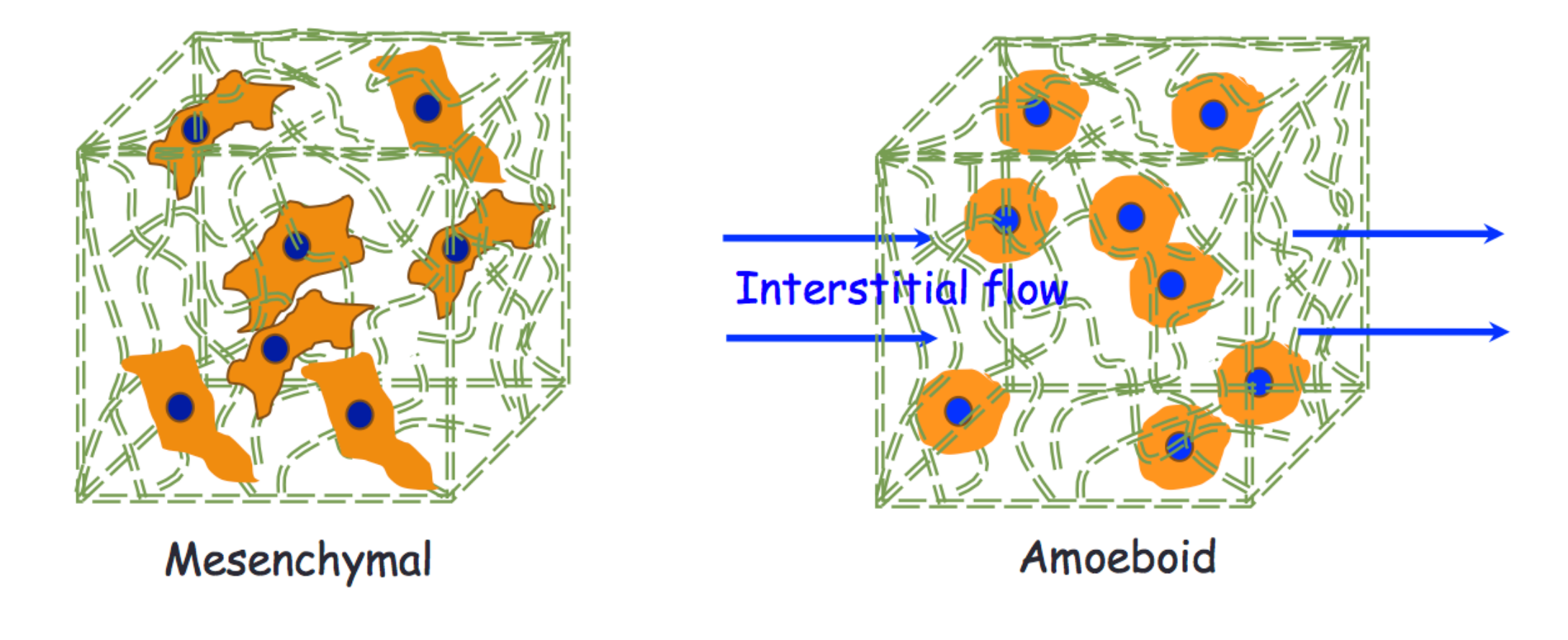}
	\vspace{-.2in}
	\caption{In the absence of flow cells exhibit the mesenchymal motility phenotype and interstitial flow promote amoeboid motility phenotype.}  
	\label{fig:rg5}
\end{figure}

Recent experiment has demonstrated that interstitial flows regulate the cancer cell migration heterogeneity within a three-dimensional biomatrix.~Using a microfluidic model, authors show that breast cancer cells (MDA-MB-231) embedded in a collagen matrix, exhibit both amoeboid and mesenchymal motility phenotype and interstitial flows
promote amoeboid over mesenchymal motility of breast cancer cells\cite{Huang15IB}.

How interstitial flows promote cancer cell invasion is largely unknown. The understanding of this process could help to develop drugs that inhibit the process and prevent cancer from metastasizing. 
How the biophysical forces modulate tumor cell migration heterogeneity and plasticity and create a complex spatiotemporal dynamical property during cancer metastasis is still elusive. 
 
In this article, we develop a theory to describe how biophysical cues regulate the diverse collective cell motility.  The cumulative effect arising from the non-equilibrium description of living cells using a time-dependent mechanical interaction, and flows, lead to complex dynamics, which may have far-reaching implications in our understanding of cancer metastasis.
One of the major difficulties in the study of collective behavior of the cells far from equilibrium is the breakdown of a fluctuation-dissipation theorem (FDT); hence, independent diagrammatic expansions for the response function and the correlation function. The equilibrium distribution is not known and averages can be computed only for the statistical noise. 

We study the relevant continuum description of the collective behavior of a colony of cells in the physical time scale, using the stochastic quantization technique, originally proposed by Parisi and Wu\cite{Parisi81ES}. We show that the time-dependent adhesion interactions that determine the structural rearrangements, the long-range hydrodynamic interactions among living cells, and self-generated force due to actin remodeling dictate the complex collective behavior, when a continuum description of the cellular colony is invoked, in the physical time scale. 
{We find that cells exhibit both amoeboid and mesenchymal motility characterized by super-diffusive motion. The mean-square displacement (MSD) of the cells for the mesenchymal motility behaves as $t^\alpha$, with, $\alpha=1.43$. The interstitial flows impair the collective migration with a gradually decreasing degree of super-diffusion and promote the amoeboid motility phenotype over mesenchymal motility. In the case of flow, the MSD exponent $\alpha=1.2$, which reflects the dynamical slowing down of the spatiotemporal collective migration. {The mesenchymal cell migration is more persistent than the amoeboid motility phenotype.
}}

{The anomalous diffusion has been observed in many contexts from single to collective motions of particles\cite{Sancho04PRL,Ramin09PRL,Romanczuk12EPJST,Metzler14PCCP, Toner16PRE}. Sancho et.al.\cite{Sancho04PRL}~shows the anomalous diffusion of particles on a solid surface controlled by the friction coefficient. In the present model, we focus on the dynamics of cells in the low Reynolds number regime, where mechanical interactions exhibit collective phenomena characterized by the super-diffusive behavior of cells. We identify the interactions based on the experiment\cite{Huang15IB} that exhibit collective behavior for the different motility phenotype. According to the experiment, the mesenchymal cells develop long-lived adhesion, we model the adhesion interaction time-dependent with a time scale of attachment $\lambda^{-1}$. We show that how the time scale of attachment is reduced by flow-induced sweeping away the FN molecules and cells exhibit amoeboid motility phenotype. We write down the density field equation for the cells. We are interested in the dynamics of cells, for example, the mean square displacement (MSD) of a cell which is given by the relation MSD$\sim t^{2/z}$, with the dynamic exponent $z$. The dynamics are out of equilibrium due to the self motility of cells and time-dependent interactions. We are interested in finite-time behavior. We use the Stochastic quantization technique to obtain the dynamic exponent, in which the correlation function is obtained in fictitious time $\tau$. The real-time correlation is obtained by taking the asymptotic limit in the fictitious time direction. The scaling exponent is obtained by simple power counting analysis. 
}

{{Mesenchymal migration is characterized by an elongated fibroblast-like morphology, highly condensed cell-matrix adhesions, and formation of contractile actomyosin bundles\cite{Pathak11IB}. 
The elongated cells demonstrated a mesenchymal phenotype where the actin filaments formed a highly polarized bundle. It is known that mesenchymal cells form long-lived adhesions with the ECM fiber bundles, which trigger the downstream signaling that activates actin remodeling and climbed along the fibers, and thus cell migration. Fibronectin (FN) is an important adhesion molecule in mediating mammalian cell migration. Cell-secreted FNs assemble into a fibrillar form and bind to collagen, which promotes a mesenchymal cell phenotype\cite{Huang15IB}. In the mesenchymal mode, cells wider than the matrix pore size extend protrusions at the leading edge to probe the surrounding ECM fibers, form stable adhesions at the poles of the elongated cell. Further polarization and strengthening of adhesions is accompanied by a rise in actomyosin contractility exerting traction force and proteolytic degradation of ECM fibers at cell-ECM junctions, and rupture the adhesions at the trailing edge of the cell, thus migrating in a ‘‘path-generating’’ manner\cite{Huang15IB,Pathak11IB}.\\ 

Amoeboid migration is characterized by a rounded morphology, formation of bleb-like protrusions, restriction of actomyosin contractility to the cell cortex, and transient, short-lived adhesions with the ECM, and squeezing through the matrix pore when finding a suitable path\cite{Huang15IB,Pathak11IB}. The inhibition of proteolysis or integrin-dependent adhesion can be compensated by a weak or non-adhesive amoeboid mode of migration in which the cell adopts a rounded morphology and changes its shape by generating hydrostatic pressure at the cell cortex, thus forcibly extruding processes through available spaces in the porous matrix and eventually deforming the cell body. In the case of flow, the interstitial-flow-induced amoeboid cell motility was likely caused by the lack of assembled endogenous adhesion molecules such as FN. More specifically, the flows carried away the cell-secreted adhesion molecules before they were assembled into fibrils and anchored to the collagen fibers\cite{Huang15IB}.} }

{\section{Theory}}
In the absence of flow, cells exhibit mesenchymal motility phenotype in 3D collagen matrix.~Cells secreted fibronectin molecules into a fibrillar form, and form long-lived adhesions with the collagen fibers, which trigger the downstream signaling that activates actin-network expansion and thus exhibit cell migration\cite{Huang15IB}.~We consider the dynamics of a colony of cells in a dissipative environment where inertial effects are negligible.
Each cell experiences systematic forces arising from mechanical interactions, and a Gaussian random force with a white noise spectrum.
The equation of motion for a single mesenchymal cell $i$ is
$
\label{eqmo}
\frac{\partial {\bf r}_i}{\partial t}=-\sum_{j=1}^{N}{\bf \nabla} U({\bf r}_i(t)-{\bf r}_j(t))+\eta_i(t)+f_0 \xi_i(t),
$
~where $U$ contains repulsive interactions with range $\lambda_1$, adhesion interactions with collagen matrix with range $\sigma_1$, and favorable attractive interactions between cells with range $\sigma$, and with strengths $v$, $g$ and $\kappa$ respectively. We use Gaussian potentials (see the appendix A1 for details) in order to obtain analytical solutions. Needless to say that the conclusions would be valid for any short-ranged  $U$.
{The highly deformable cells with flexible cell boundaries connected through cell-cell junctions, which can be ruptured and reconnected. Cell structural rearrangement for changing nearest-neighbor cells has been seen during the migration of cells in movie S1 in \cite{Huang15IB}. Therefore we model the cell-cell interaction as time-dependent potential.} We assume that the adhesion strength is changing during the topological rearrangement via $a_f + (a_i -a_f)e^{-\lambda t}$\cite{Himadri04PRE}. $a_i $ and $a_f$ are initial and final interaction strengths ($a$ stands for $g$ and $\kappa$) and the time scale for changing the receptor-ligand interaction is given by $\lambda^{-1}$. {{In addition to short-range interactions, the cells that are well separated, interact with each other via the matrix. The cells in the collagen matrix can be modeled as force concentration dipoles. The cells interact via long-ranged elastic interaction potentials ($\sim 1/r^3$ with distance $r$)\cite{Schwarz13RMP,SOPHER181BJ,Schwarz02PRL}. The interaction potentials depend on elastic constants, geometry, and cellular orientations.  For simplicity of the calculation, we consider the interaction is isotropic and repulsive. The potential $U_{elastic} = \frac{(2+\Lambda)^2P^2}{4\pi c(1+\Lambda)} (1/r^3)$}, where $P$ is the magnitude of the force dipole, $\Lambda=\lambda/\mu$, and $c=2\mu+\lambda$, where $\mu$ and $\lambda$ are the Lam$\acute{\text{e}}$ coefficients of the isotropic elastic medium\cite{Schwarz02PRL}.} The Gaussian white noise, satisfies $<\eta_i(t)\eta_j (t')>=2 D\delta_{ij}\delta(t-t')$. The mesenchymal cells are subject to a self-generated force of actin-network remodeling. The cells are, thus in addition subject to a random self-generated force with amplitude $f_0$. The randomness is modeled by an athermal noise $\xi(t)$, which is exponentially correlated over a time scale $\tau_p$. The statistics of the $\xi(t)$ is given by $<\xi(t)>=0$, $<\eta(t)\xi(t')>=0$ and $\xi(t)\xi(t')=b \exp[-| t-t' | / \tau_p]$. Where $b$ is the dimensionless constant. The athermal noise in general does not obey the FDT.

{For the mesenchymal phenotype, the cells form long-lived adhesion with the matrix and exhibit matrix mediated long-range elastic interactions when they are well separated. 
On the other hand, the amoeboid cells form short-lived adhesion with the matrix and migrate squeezing through the pore. Therefore, we assume amoeboid cells do not feel long-range elastic interactions in the long-time limit. 
}

{ In the absence of flow, cell-secreted FNs assemble into a fibrillar form and bind to collagen, which promotes a mesenchymal cell phenotype. In the case of flow, the interstitial-flow carried away the cell-secreted adhesion molecules before they were assembled into fibrils and anchored to the collagen fibers and thus amoeboid cell motility was likely caused by the lack of assembled endogenous adhesion molecules such as FN. Therefore, in the case of flow cells form short-lived adhesion with matrix and form rounded morphology and thus amoeboid motility\cite{Huang15IB}. The experiment in \cite{Huang15IB} investigates the role of exogenous FN in cell morphology and motility in the absence/presence of the flows. 
{In the presence of flow how the time scale of adhesion is reduced by sweeping away the cell-secreted adhesion molecules and therefore the cells exhibit amoeboid motility, can be understood in the following way.
We write the concentration equation for the FN molecules which is advected by the flow,
$\frac{\partial C}{\partial t}+\nabla \cdot{(C {v})}=-\lambda C$. Where $\lambda $ is the rate of FN molecules assembled into fibrilar form and the fluid velocity follows, $\eta \nabla^2 v=\nabla \cdot f$, with Gaussian random noise $f$. Assume the strength $a$ depends on the concentration of FN molecules. We write $a(C)= a_1 + a_2 \frac{\delta C}{C_0} + a_3 (\delta C/C_0)^2$. From the concentration equation for FN molecules and velocity equation, we obtain the average adhesion strength $<a>= a_1 +a_3(1-\exp[-2\lambda t])$. The average adhesion strength can be written in the form $a_f +(a_0-a_f)\exp[\lambda' t]$, where, $a_1=a_0$ and $a_3 =a_1 +a_f$. Now the time scale of adhesion, i.e, $(\lambda')^{-1} =\frac{1}{2\lambda}$ is reduced by factor of 2. The cell becomes shortlived and exhibit ameboid motility phenotype. In the presence of flow we use the time scale of adhesion is $(\lambda')^{-1}$.}
The amoeboid cells migrate through squeezing the matrix pore when finding a suitable path.  Therefore we consider the long-range hydrodynamic interactions between amoeboid cells leading to collective behavior of cell motility.}
We begin by considering the dynamics of a colony of cells in the presence of flow in a dissipative environment where inertial effects are negligible.
Each cell experiences mechanical forces, such as adhesion, excluded volume interactions due to neighbors, and a random force characterized by Gaussian white noise. 
The equation of motion for single cell $i$ is~\cite{Ermak78JCP}
\begin{eqnarray}
\label{eqmo}
\frac{\partial {\bf r}_i}{\partial t}&=&k_B T\sum_{j=1}^{N}   \mu_{ij}  {\bf \nabla}_{r_j} U({\bf r}_i(t),{\bf r}_j(t))\\ \nonumber
&&
+\eta_i(t)+f_0 \xi_i(t).
\end{eqnarray}
The first term on the r.h.s.~of Eq.(\ref{eqmo}) is the effect of force acting on cell $j$ creates a hydrodynamic flow-field in the fluid, thereby entraining cell $i$.
Where $U$ contains repulsive interactions with range $\lambda_1$, adhesion interactions with collagen matrix with range $\sigma_1$, and favorable attractive interactions between cells with range $\sigma$, and with strengths $v$, $g$, and $\kappa$ respectively (see the appendix for details).

$\eta_i$ is assumed to be Gaussian random vectors exhibiting hydrodynamic correlations according to fluctuation-dissipation theorem (FDT),
 $<\eta_i(t)\eta_j (t')>=2 k_B T D~\overleftrightarrow{\mu}_{ij}\delta(t-t')$.~{The dynamics of the cell is described by $\frac{\partial r}{\partial t}=u(r)$, where $r$ is the position of the cell. The flow velocity of the fluid at $r$ is determined by the solution of the Stokes equation, $\eta \nabla^2 u(r)=\nabla P+F$, where, $F=\sum_i^{N} f \delta(r)$. We assume that the fluid is incompressible, i.e., $\nabla \cdot u=0$. From these set of equations, the Stokes equation is readily solved to obtain a fluid velocity at the center of the cell at ${\bf r}$ for a point force ${\bf f}$ as, {\bf u(r)}=$k_B T\sum_{j=1}^{N}   \mu_{ij} f_j$, where $ \mu_{ij}=\frac{1}{8\pi \eta r}(\delta_{i j}+\frac{r_i r_{j}}{r^2})$ is the mobility matrix with the relative coordinate between cell $i$ and $j$, ${\bf r}={\bf r}_i -{\bf r}_j$. The long-ranged nature of hydrodynamic interaction apparent from ${\bf u(r)}$ suggests that a collection of cells that dynamically exert forces on the fluid medium they are immersed in could influence each other very strongly, leading to the possibility of collective behaviors.}
 The protrusive flowing of the anterior actin network of the cell and squeezing actomyosin contractions of the trailing edge are modeled as the cells are subjected to a self-generated force with amplitude $f_0$ during their pathfinding motion in the collagen matrix. The statistics of the $\xi(t)$ is the same as for mesenchymal motility.


 We consider the evolution of the density function for a single cell $\phi_i({\bf r},t)=\delta[\bf r-{\bf r}_i(t)]$. A closed form Langevin equation for the density, $\phi({\bf r},t)=\sum_i \delta[\bf r-{\bf r}_i(t)]$ can be obtained using standard approach~\cite{Dean96JPA}. The time evolution of $\phi({\bf r},t)$ is given by,
\begin{eqnarray}\label{phi11}
&&\frac{\partial \phi({\bf r},t)}{\partial t}= {\bf \nabla }\cdot \left(\phi({\bf r},t)\int d{\bf r'} \phi({\bf r'},t)\overleftrightarrow{\mu}{\bf \nabla}U({\bf r-\bf{r'}})\right)\\ \nonumber &&+D \nabla^2 \overleftrightarrow{\mu}\phi({\bf r},t) +{\bf \nabla} \cdot \left((\eta+\xi) \phi^{1/2}({\bf r},t)\right).
\end{eqnarray}
~Note that the density equation contains the same information as N-body stochastic Langevin equations.
 This is an out of equilibrium problem characterized by the absence of fluctuation-dissipation theorem due to long-range hydrodynamic term and self-generated force due to actin remodeling.
Eq.~(\ref{phi11}) can be studied analytically by treating the non-linear terms using a perturbative approach, based on the stochastic quantization scheme~\cite{Parisi81ES,Himadri06PLA,Himadri06PRE}.

{\subsection{Stochastic quantization approach}}
To understand the dynamics of collections of cells, we use the stochastic quantization method developed by Parisi-Wu in another context. 
The collective migration of cells described by Eq.~(\ref{phi11}) is an out of equilibrium problem characterized by the absence of FDT, which relates the correlation and response function in momentum space as $C=\frac{1}{w} \text{Im} G$. 
The usual analytic route to get the scaling solution of this problem, one can introduce a response field $\tilde{\phi}$. We need to calculate both the response function ($G=<\phi \tilde{\phi}>$) and correlation function ($C=<\phi \phi>$) because of the 
absence of a fluctuation-dissipation relation.
The key advantage of the present method is that we do not need to calculate both the correlation and response functions. The FDT is constructed in fictitious time introduced in the problem. The FDT relation enables us to obtain the scaling of the correlation function, once the scaling of the response function is known. By taking the infinite limit in fictitious time, one can obtain the correlation function in real-time. The scaling solution of the problem can be obtained by power counting analysis instead of doing renormalization group calculation.

We now exploit the Parisi-Wu stochastic quantization scheme~\cite{Parisi81ES}, and introduce a fictitious time $`\tau_f$', and 
consider all variables to be functions of $\tau_f$ in addition to {\bf k} and $w$. 
A Langevin equation in $`\tau_f$' space is,
\begin{equation}\label{langefic1}
\frac{\partial \phi_1({\bf k},w,\tau_f)}{\partial \tau_f}=-\frac{\delta \mathcal{S}}{\delta \phi_1(-{\bf k},-w,\tau_f)}+f_{\phi_1}({\bf k},w,\tau_f) \, ,
\end{equation}
with $<f_{\phi_1} f_{\phi_1}>=2 \delta(k+k')\delta(w+w')\delta(\tau_f-\tau_f')$.
This ensures that as $\tau_f\rightarrow \infty$, the distribution function will be given by the action $S({\bf k},w)$, because in the $\tau_f$-space  a fluctuation dissipation theorem (FDT) is preserved. The action $S({\bf k},w)$ can be obtained by writing down the probability distribution $P(f_{\phi_1}) \propto \text{exp}[-\int_{{\bf k},w} \{\frac{1}{2(Dk^2\mu(k)\phi_0+(f_0^2 k^2 \xi(\omega) )\phi_0)}f_{\phi_1}({\bf k},w)f_{\phi_1}(-{\bf k},-w) ]\\=\text{exp}[-\mathcal{S}]$ corresponding to the noise term $f_{\phi_1}$, and  the action $S({\bf k},w)$ in terms of $\phi_1({\bf k},w)$ with the help of Eq.(\ref{phi11}), {where $\mu(k)=\frac{1}{8\pi\eta}(\frac{\delta_{ij}}{k^2}-\frac{k_i k_j}{k^4})$.} The expression for the $\mathcal{S}$ is in the appendix.

\begin{figure}[b]
	\includegraphics[width=.450\textwidth]{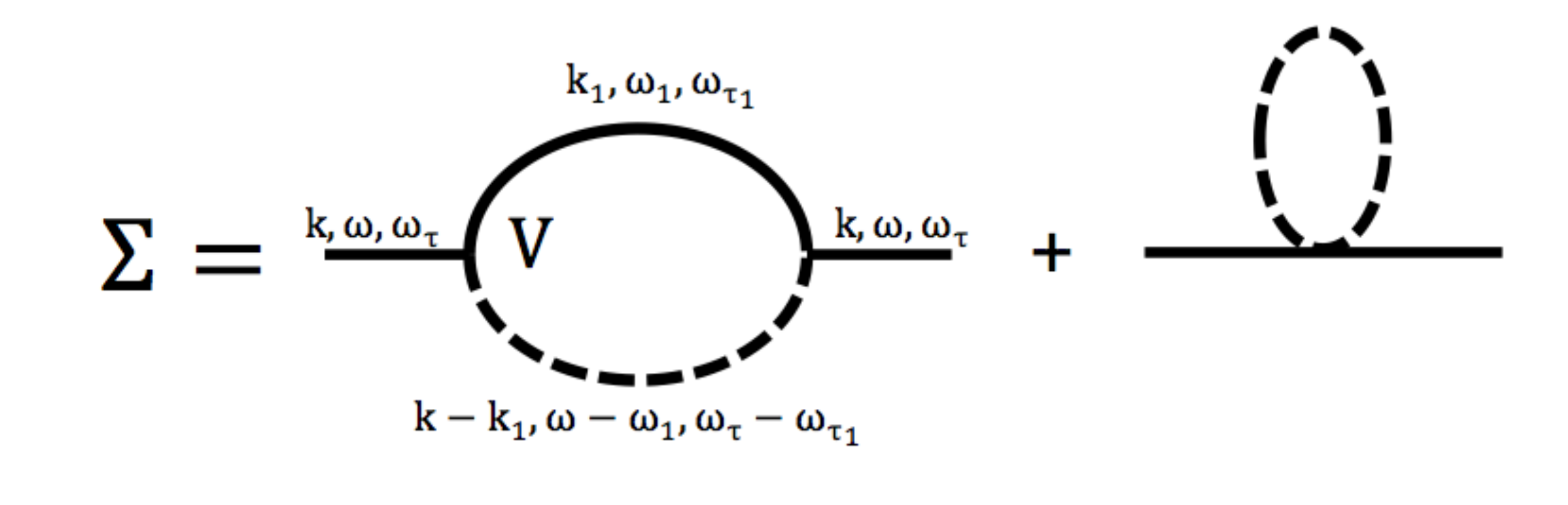}
	\caption{Dashed line indicates the correlation function ($G_0 G_0^*$) and solid line indicates the response function ($G_0$).  Self-energy term ($\Sigma$) is obtained by contracting the two $\phi_1$ fields. First term is the two loop contribution from the first order term (contains two $\phi_1$ fields) in the fictitious time equation. Second one is the one loop contribution from second order term (contains three $\phi_1$ fields). }  
	\label{fig:rg6}
\end{figure}

We follow the procedure of obtaining scaling laws of the problem, which has been demonstrated in earlier works~\cite{Himadri18PRE,Himadri06PRE, Himadri06PLA,Abdul18PRX}. The dynamics of Eq.(\ref{langefic1}) requires only the calculation of response functions, the correlation functions in this dynamics are related to response function through FDT relation, i.e., in Fourier space, $C=\frac{1}{\omega_\tau}\text{Im} G$. We can obtain the scaling laws in real space and time in a straight forward fashion from the solution in the fictitious time $'\tau'$ space.  

We obtain the following self-consistent equation for the self-energy from the calculation of response function using Eq.(\ref{langefic1}):
\begin{equation}\label{scale}
\Delta\nu =\frac{D_0}{2\nu } \Sigma({k},\omega, \omega_{\tau_f}),
\end{equation}
where, $\nu=D\mu(k) k^2+\phi_0 k^2 \mu(k)g(\omega)U({\bf k})$, $D_0=2(Dk^2\mu(k)\phi_0+(f_0^2 k^2 \xi(\omega) )\phi_0)$, and $\Sigma$ is the self-energy term, a two-loop contribution from the first order term ( containing two $\phi_1$ fields)  in Eq.~(\ref{langefic1}) (first term in Fig.~\ref{fig:rg6}) will contribute in the scaling laws for the cell in the finite time. We use Eq.(\ref{scale}) for getting the scaling laws of both the amoeboid and mesenchymal cells phenotype.

{\section {Results}} 
{\subsection{Mesenchymal motility}} 
The mesenchymal cell phenotype forms long-lived integrin-based adhesions with the collagen matrix and migrate via either remodeling of actin network or degrading the matrix. The non-linear term i.e. the adhesion interaction plays an important role in the complex dynamics of collective migration of mesenchymal cell phenotype.
In a self consistent mode coupling theory, we now replace $\nu$ by $\Delta \nu$ in 
the self energy term $\Sigma({0},\omega, \omega_{\tau_f})$ in the first term in Fig.(\ref{fig:rg6}), use $G\sim \omega_{\tau_f}^{-1}$  as from 
Eq.~(\ref{langefic1}), and $C$, which follows from the FDT. In the absence of flow, $\mu=1$.
{According to scale transformation, we know that $\omega \sim k^z$, $\omega_\tau \sim k^{4z-2}$, $G \sim k^{-4z+2}$, $C \sim k^{-8z+4}$ and the vertex factor $V \sim k^{2z}$. The self energy term in Fig.(\ref{fig:rg6}) can be written as 
$\Sigma({0},\omega, \omega_{\tau_f})\sim  \int \frac{d^d {\bf k'}}{(2\pi)^d} \frac{d\omega'}{2\pi} \frac{d\omega'_\tau}{2\pi} V V GC$.}
By carrying out the momentum count of $\Sigma({0},\omega, \omega_{\tau_f})$, and 
using $\Delta \nu \sim k^z$, we find that $\Sigma({\bf k},\omega, \omega_{\tau_f})\sim k^{d+4-3z}$. 
Using Eq. (\ref{scale}) and $\nu/D_0 \sim k^{z}$, we have $k^{2z}\sim k^{d+4-3z}$, which leads to $z=\frac{d+4}{5}$.
MSD exponent $\alpha=2/z=10/(d+4)$. In 3D, $\alpha=1.43$, i.e., the mesenchymal cells undergo super-diffusion. 
The non-linear term arising from cell-cell adhesion that determines the dynamical shape change during cell motion and self-generated force, produce super-diffusive motion. {The theoretical result is in good agreement with the recent experimental result ($1.46\pm 0.013$) using the microfluidic model\cite{Huang15IB}. If we wait longer than experimental observation time, the cells become apart from each other. The short-range interactions do not play any role in cell motility. The long-range elastic interactions dictate the dynamics of the mesenchymal phenotype. The similar scaling analysis shows the super-diffusive behavior of cells characterized by the single-cell MSD exponent $\alpha=1.15$. The degree of superdiffusion decreases in the long-time where long-range elastic interactions play a role in cell motility.
}  

\begin{figure}[h]
 \centering
   \subfloat[]{{\includegraphics[width=7cm]{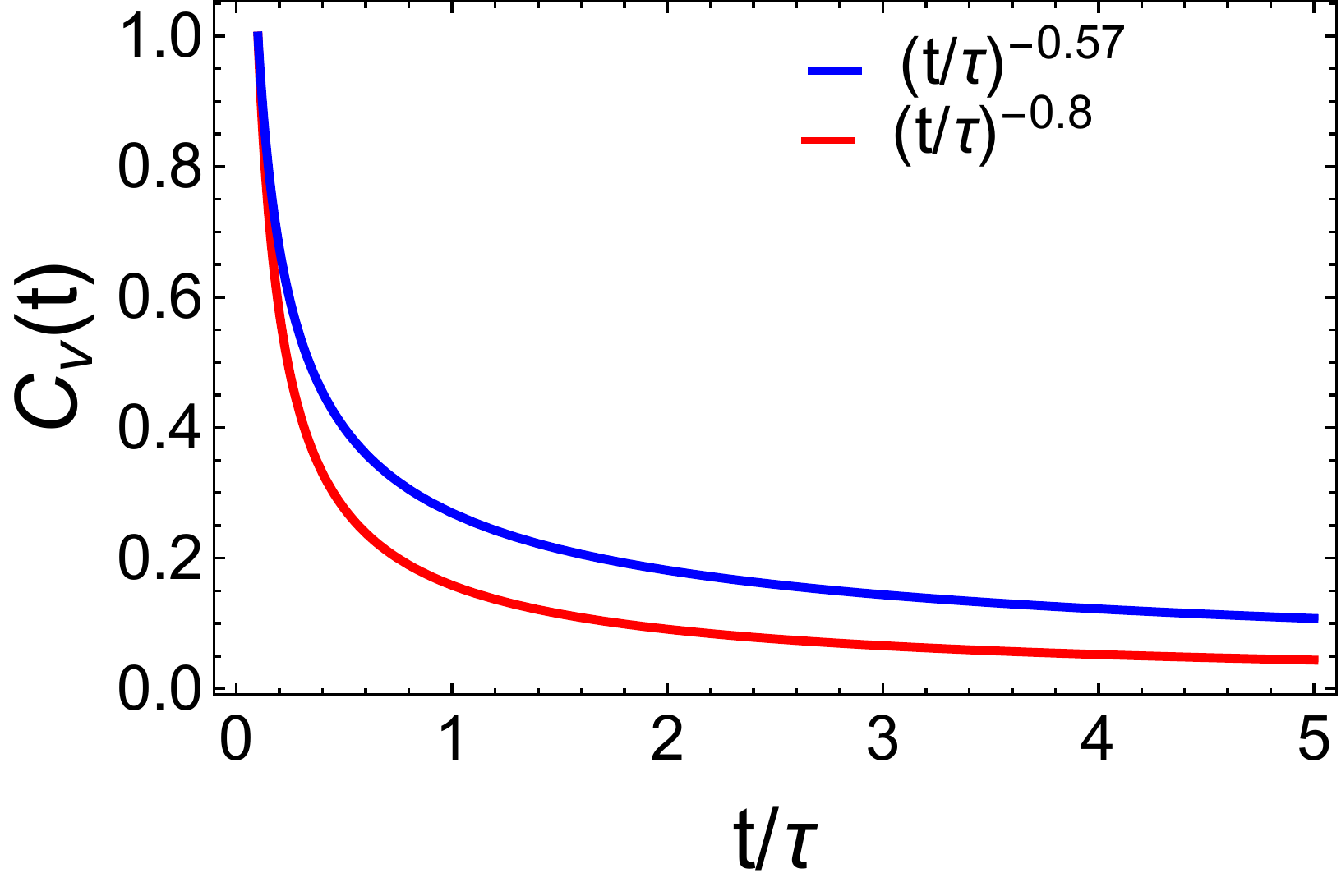} }}%
    \qquad
    \subfloat[]{{\includegraphics[width=6.7cm]{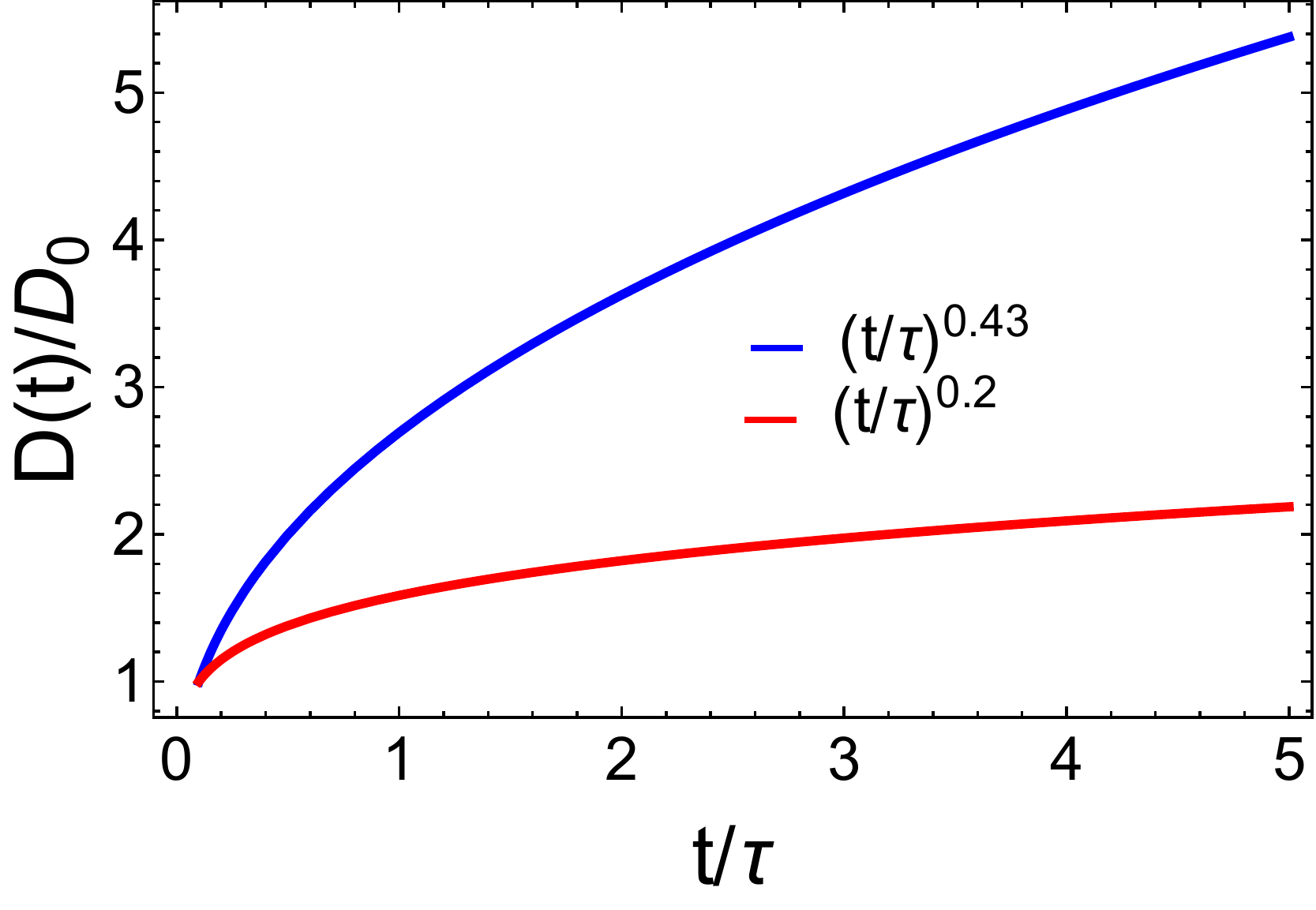} }}%
    \label{fig:rg54}%
	\caption{(a) Velocity auto-correlation function, normalized to unity at $t/\tau=0.1$, where $\tau=\nu^{-1}$. (b) The blue and red lines dictate the diffusivities ($D(t)$) for mesenchymal and amoeboid cell motility respectively. } 
\end{figure}
{\subsection {Interstitial flow induced amoeboid motility}}
In the presence of interstitial flow, cells exhibit amoeboid motility. The amoeboid cells form short-lived adhesion with the collagen matrix. The time scale $\lambda'^{-1}$ is small compared to the mesenchymal motility phenotype.  
In the case of flow, the cells exhibit long-range hydrodynamic interactions that determine the complex spatiotemporal dynamics of amoeboid cell phenotype. The self-generating force of actin remodeling helps to propel in pathfinding fashion through the collagen matrix.
In a self consistent mode coupling theory, we now replace $\nu$ by $\Delta \nu$ in 
the self energy term $\Sigma({0},\omega, \omega_{\tau_f})$ in the first term in Fig.(\ref{fig:rg6}), use $G\sim \omega_{\tau_f}^{-1}$  as from 
Eq.~(\ref{langefic1}) and $C$, which follows from the FDT. 
{According to scale transformation, we know that $\omega \sim k^z$, $\omega_\tau \sim k^{4z-2}$, $G \sim k^{-4z+2}$, $C \sim k^{-8z+4}$ and the vertex factor $V \sim k^{4z-2}$. The self energy term in Fig.(\ref{fig:rg6}) can be written as 
$\Sigma({0},\omega, \omega_{\tau_f})\sim  \int \frac{d^d {\bf k'}}{(2\pi)^d} \frac{d\omega'}{2\pi} \frac{d\omega'_\tau}{2\pi} V V GC$.}
By carrying out the momentum count of $\Sigma({0},\omega, \omega_{\tau_f})$, and 
using $\Delta \nu \sim k^z$, we find that $\Sigma({\bf k},\omega, \omega_{\tau_f})\sim k^{d+z}$. 
Using Eq. (\ref{scale}) and $\nu/D_0 \sim k^{3z-2}$, we have $k^{4z-2}\sim k^{d+z}$, which leads to $z=\frac{d+2}{3}$.
MSD exponent $\alpha=2/z=6/(d+2)$. In 3D, $\alpha=1.2$, i.e., the amoeboid cells undergo superdiffusion. 
The non-linear term due to long-range hydrodynamic interactions among cells besides self-generated force, produce super-diffusive motion for amoeboid cell phenotype. The decrease of MSD exponent determines the dynamical slowing down of initial collective migration of the mesenchymal cell phenotype. Therefore, interstitial flows impair the cell's ability to spread by sweeping away the adhesion molecules with the flow and making the cells as amoeboid phenotype with short-lived adhesion with the collagen fiber\cite{Huang15IB}. The cells migrate via squeezing through the pore of the collagen fiber when finding a suitable path\cite{Pathak11IB}. {The theoretical result is in good agreement with the recent experimental result ($\alpha=1.27 \pm 0.013$) using the microfluidic model\cite{Huang15IB}. If we wait longer than experimental observation time, the cells become apart from each other. The short-range interactions do not play any role in cell motility. The cells undergo normal diffusion for the amoeboid motility phenotype characterized by the single-cell MSD exponent $\alpha=1.0$.
}

{We calculate velocity auto-correlation function $C_v(t)=\frac{1}{2d}\frac{d^2}{dt^2}<\delta r^2(t)>$, and time-dependent diffusion coefficient, $D(t)=\frac{1}{2d}\frac{d}{dt}<\delta r^2(t)>$, where $<\delta r^2(t)>$ is the single cell MSD. The blue line dictates the mesenchymal motility and red line shows the amoeboid motility. The slower decay of velocity correlation function shows that the mesenchymal cells are more persistent than amoeboid cells. 
Similarly, the time-dependent diffusion coefficient for a mesenchymal cell is higher than the amoeboid cells, i.e, the mesenchymal cell is more diffusive than amoeboid cells. The theoretical result is in agreement with the experiment\cite{Huang15IB}.}

{The MSD for a single cell in the experiment in \cite{Huang15IB} shows the super-diffusion with the exponent $\alpha>1$ for both with and without the flow. 
Average for MSD calculation has been taken over all cell trajectories (approximate over 60 motile cells).  The movies (S1 and S2) in \cite{Huang15IB} show that cells are interacting through adhesion interactions that influence each other very strongly leading to the possibility of collective behaviors. 
That implies that the super-diffusion arises due to the collective motion of interacting cells for both with and without flow. 
Although it is a single-cell measurement, the super-diffusive motion of a cell arises due to the collective migration of interacting cells.\\ 
In theory, the flow effect that we model through long-range hydrodynamic interactions among cells. That induces the super-diffusion for motile amoeboid cells. 
On the other hand, the time-dependent short-range interactions with the coupling with self-motility produce super-diffusive motion in mesenchymal cells. 
The single-cell MSD shows super-diffusion.  The time-dependent short-range interactions and long-range hydrodynamic interactions between cells influence each other very strongly and produce non-linear terms in the density equations, leading to the possibility of novel collective behaviors.  
Therefore the super-diffusion effect is a collective phenomenon.
}

{\section{Conclusion}}
In the present contribution, using a new theoretical framework, we provide insight into the dynamics of a colony of cancer cells driven by biophysical cues. The theory reveals that the interstitial flows regulate cancer cell morphology and motility phenotypes, emphasizing the role of fluid flows in regulating cancer cell migration heterogeneity. The conventional practice in dealing with this out of equilibrium problems is to use a set of fictitious fields called response fields, which provide a field-theoretic prescription for the response function. In contrast, we propose the introduction of a fictitious time in which an FDT is valid, and thereby only correlation functions need to be calculated. Our approach greatly simplifies the evaluation of scaling exponents. 
We find that the non-linear term in the density evolution equation arising from mechanical interactions along with self-generating force due to actin remodeling determines the scaling behavior for the collective migration of cells. In the absence of flow, cells exhibit collective migration of mesenchymal motility phenotype induced by time-dependent interaction potentials that determine the structural rearrangement during their path generating migration through the collagen matrix. In contrast, the cells exhibit the amoeboid motility in the presence of flow and exhibit a dynamical slowing down of directed migration, with a gradually decreasing degree of super-diffusion. The long-range hydrodynamic interactions among cells in the presence of interstitial flow determine the collective migration of cells through the matrix pore in a pathfinding fashion. {The mesenchymal cells are more persistent and diffusive compared to amoeboid cells.}
The theoretical framework introduced here provides evidence of interstitial flow directed collective motion heterogeneity and could explain the invasion of cancer cells under interstitial flow, observed in a recent experiment\cite{Huang15IB}. The theory introduced here could help us understand how cancer cells spread by invading adjacent tissues involved in metastasis~\cite{Polyak09NRC}.
\section{acknowledgements}

I wish to express my gratitude to Prof. D. Thirumalai for supporting me as a postdoc at the department of chemistry at UT, Austin. I am indebted to Sumit Sinha and Dr. Debayan Chakraborty for valuable comments.

\appendix
\section{Short-range interaction}
To obtain the dynamics of an evolving collection of cells, we use the following simplified form for cell-cell interaction,
\begin{equation}\label{HamiltVH}
{U}({\bf r}(i)-{\bf r}(j))=\frac{v}{(2\pi \lambda_1^2)^{3/2}} 
e^{-\frac{({\bf r}(i)-{\bf r}(j))^2}{2\lambda_1^2}}-
\frac{\kappa}{(2\pi \sigma^2)^{3/2}} e^{-\frac{({\bf r}(i)-{\bf r}(j))^2}{2\sigma^2}},
\end{equation}
where $v$ and $\kappa$ are the strengths of excluded volume and attractive interactions, respectively.

In addition, the cell surface-ECM interactions $U_{s}$, determining the configuration-dependent forces experienced by the cells: $U_{s}=-\frac{g}{(2\pi \sigma_1^2)^{3/2}} e^{-\frac{({\bf r}_i-{\bf r}_0)^2}{2\sigma_1^2}}$. {The potential term $U_s$ describes the cell surface-collagen interactions as a function of $r_0$, the average distance between cell and collagen. Cells interact with the collagen through receptor-ligand interactions, described by short range potential.
Mesenchymal cells form long-lived adhesion with collagen. We assume the adhesion strength is changing during the topological rearrangement via, $a_f + (a_i -a_f)e^{-\lambda t}$. $a_i $ and $a_f$ are initial and final interaction strengths and the time scale of changing the receptor-ligand interaction is given by $\lambda^{-1}$. Where $a$ stands for $g$ and $\kappa$. In contrast, amoeboid cells form short lived adhesion with the collagen, i.e., the time scale for the adhesive interaction $\lambda^{-1}$, is small compared to mesenchymal cells.

\section{Density equation}
To simplify the multiplicative noise term (last term in Eq.~(2) in the main text), we assume that the density fluctuates around a constant value.
Hence, we define the density using $\phi({\bf r},t)=\phi_0+\phi_1({\bf r},t)$, and expand Eq.(1) in the main text in $\frac{\phi_1}{\phi_0}$ up to the lowest order in non-linearity. 
In Fourier space, the equation for the density fluctuation becomes, 
\begin{eqnarray}
\label{rho12}
&&\frac{\partial \phi_1({\bf k},t)}{\partial t}=-((Dk^2 +\phi_0 k^2 a(\omega) U({\bf k}))\mu(k))\phi_1({\bf k}) \\ \nonumber&& 
+\int d{\bf q} (-{\bf q}\cdot {\bf k})\mu({\bf q})U({\bf q}) a(\omega)\phi_1({\bf q})\phi_1({\bf k}-{\bf q})+
f_{\phi_1},
\end{eqnarray}
with $<f_{\phi_1}f_{\phi_1}>=Dk^2\mu(k)\phi_0+k^2 \xi(\omega)\phi_0$. Where, $\xi(\omega)=\frac{2\tau_p}{1+\tau_p^2 \omega^2}$.
The Greens function $G$ is given by,
\begin{equation}\label{green1}
[G]^{-1}=-i \omega+Dk^2 \mu(k)+\phi_0 k^2 a(\omega) \mu(k)U({\bf k})+\Sigma({\bf k},\omega) ,
\end{equation}
where, $a(\omega)=\frac{1}{\lambda +i \omega}$, and $\Sigma({\bf k},\omega)$ is the self energy term contributed from non-linear adhesion interactions.

\section{ Stochastic quantization technique} We now exploit the Parisi-Wu stochastic quantization scheme~\cite{Parisi81ES, Himadri06PRE, Himadri06PLA, Himadri18PRE}, and introduce a fictitious time $`\tau_f$', and 
consider all the variables to be functions of $\tau_f$. 
A Langevin equation in $`\tau_f$' space is,
\begin{equation}\label{langefic}
\frac{\partial \phi_1({\bf k},w,\tau_f)}{\partial \tau_f}=-\frac{\delta \mathcal{S}}{\delta \phi_1(-{\bf k},-w,\tau_f)}+f_{\phi_1}({\bf k},w,\tau_f) \, ,
\end{equation}
with $<f_{\phi_1} f_{\phi_1}>=2 \delta(k+k')\delta(w+w')\delta(\tau_f-\tau_f')$.
Because FDT is valid in the fictitious time it follows that as $\tau_f\rightarrow \infty$, the distribution function will be given by the action $S({\bf k},w)$. The action $S({\bf k},w)$ can be obtained by writing down the probability distribution $P(f_{\phi_1}) \propto \text{exp}[-\int \frac{d^d{\bf k}}{(2\pi)^d}\frac{dw}{2\pi}\{\frac{1}{2(Dk^2\mu(k)\phi_0+(f_0^2 k^2 \xi(\omega) )\phi_0)}f_{\phi_1}({\bf k},w)\\f_{\phi_1}(-{\bf k},-w) ]=\text{exp}[-\mathcal{S}]$ corresponding to the noise term $f_{\phi_1}$ in Eq.(\ref{rho12}), and  the action $S({\bf k},w)$ in terms of $\phi_1({\bf k},w)$ using Eq.(\ref{rho12}). The expression for the action $\mathcal{S}$ obtained using Eq.(\ref{rho12}) is,

 
$\mathcal{S}=\int \frac{d^d{\bf k}}{(2\pi)^d}\frac{dw}{2\pi}\frac{1}{2(Dk^2\mu(k)\phi_0+(f_0^2 k^2 \xi(\omega) )\phi_0)}
\{-i\omega+(Dk^2 \mu(k)+\phi_0 k^2 a(\omega) \mu(k)U({\bf k})) \phi_1({\bf k})
+\int d{\bf q} (-{\bf q}\cdot {\bf k})\mu({\bf q})a(\omega)U({\bf q})\phi_1({\bf q})\phi_1({\bf k}-{\bf q})\}
\{i\omega+(Dk^2 \mu(-k)+\phi_0 k^2 \mu(-k)U({-\bf k})) \phi_1({-\bf k})
+\int d{\bf q} (-{\bf q}\cdot {-\bf k})\mu({\bf q})a(\omega)U({\bf q})\\ \phi_1({\bf q})\phi_1({-\bf k}-{\bf q})\}$
\section{Greens function for density equation}
The correlation functions, calculated from Eq.~(\ref{langefic}), lead to the required correlation functions of the original theory, in the $\tau_f \rightarrow \infty$ limit. In order to obtain the scaling laws, it suffices to work at arbitrary $\tau_f$. It is  obvious from Eq.~(\ref{langefic}) that in the absence of the non-linear terms, the Greens function $G^{(0)}$ is given by,
\begin{eqnarray}\label{green}
&&[G^{(0)}]^{-1}=-i\omega_{\tau_f}+\frac{1}{2(Dk^2\mu(k)\phi_0+(f_0^2 k^2 \xi(\omega) )\phi_0)}\\ \nonumber
&&[  \omega^2 +\{ D\mu(k) k^2+\phi_0 k^2 \mu(k)a(\omega)U({\bf k})\}^2] \, ,
\end{eqnarray}
~where $\omega_{\tau_f}$ is the frequency corresponding to the fictitious time $\tau_f$. As is customary, the effect of non-linear terms, can be included perturbatively leading to the Dyson's' equation
\begin{equation}\label{green1}
[G]^{-1}=[G^{(0)}]^{-1}+\Sigma({\bf k},\omega, \omega_{\tau_f}) 
\end{equation}
Here, we are concerned with the behavior of $\Sigma({\bf k},\omega, \omega_{\tau_f})$, which becomes non-linear when expanded to second order. We note that the contribution comes from two different sources (1) 
a one-loop contribution from the second order term (containing three $\phi_1$ fields) in Eq.~(\ref{langefic}) (second term in Fig.\ref{fig:rg6}) and (2) a two-loop contribution from the first order term ( containing two $\phi_1$ fields)  in Eq.~(\ref{langefic}) (first term in Fig.~\ref{fig:rg6}).  The  contribution arising coming from the term containing three $\phi_1$ fields, in Eq.~(\ref{langefic}) can 
be readily obtained by contracting two of the $\phi_1 $ fields. The  second order term coming  from the one loop contribution in 
Eq.~(\ref{langefic}) does not have any new momentum dependance. Hence it is the second-order contribution (first term in Fig.(\ref{fig:rg6})), coming from the two-loop contribution in Eq.~(\ref{langefic}), which is  significant.
The correlation function is given by the FDT as $C=\frac{1}{\omega_{\tau_f}} \text{Im}G$.  
With these observations, Eq. (\ref{green1}) can be written as,
\begin{equation}\label{green2}
[G]^{-1}({\bf k},\omega,\omega_{\tau_f})=-i\omega_{\tau_f}+\frac{1}{2(D_0)}[  \omega^2 ]+\frac{1}{2(\bar{D})}[\nu_{eff}^2 ] \, ,
\end{equation}
where $D_0=2(Dk^2\mu(k)\phi_0+(f_0^2 k^2 \xi(\omega) )\phi_0)))$, and $\bar{D}$ is defined by
\begin{equation}\label{green3}
\frac{1}{2(\bar{D})}[ \nu_{eff}^2 ]=\frac{1}{2(D_0)}(\nu  )^2+\Sigma({\bf k},\omega, \omega_{\tau_f})
\end{equation}
with $\nu=D\mu(k) k^2+\phi_0 k^2 \mu(k)a(\omega)U({\bf k})+k^2 \mu(k)$. 
Expanding $\nu_{eff}$, $\bar{D}$ about $\nu$ and $D_0$, respectively, and noting that the renormalization of $\nu $ dominates, we get
\begin{eqnarray}\label{scale}
&&\nu_{eff} \simeq \nu  +\frac{D_0}{2\nu } \Sigma({0},\omega, \omega_{\tau_f}),\\ \nonumber ~~\text{or},~&&\Delta\nu =\frac{D_0}{2\nu } \Sigma({0},\omega, \omega_{\tau_f})
\end{eqnarray}

{Collagen gels are used extensively for studying cell-matrix mechanical interactions. The cells are interacting with ECM. The Reynolds number is small (an estimate:
density $\rho=2.7 Kg/m^3$, viscosity $\mu=6.6 \times 10^4 Kg/m.s$, L=$10^{-5} m$, velocity $u=10\mu m/hr$, the Reynolds number $R=\frac{\rho L u}{\mu}=10^{-18}$), an overdamped approximation is appropriate implying that the neglect of the inertial term $m \ddot{r}\approx 0$ is
justified. Since additional adhesive forces are also present, cell movement is further damped.
The friction is high (an estimate:$\mu=6.6 \times 10^4 Kg/m.s$, radius R=$5 \times 10^{-6} m$, then friction $\gamma =6 \pi \mu R= 6.2 Kg/s$), compared to water ($\gamma=8.38 \times 10^{-8} Kg/s$, for viscosity $\mu=8.90 \times 10^{-4} Kg/m.s$). By changing the viscosity we can change the friction and hence the diffusion coefficient ($D=k_B T/\gamma$). The cells are self motile. The estimate for the diffusion due to self motility: $D_{sp}=\frac{<r^2>}{2 d \tau}=3 \times 10^{-4}m^2/s$ (where the values of $<r^2>=10^{-8}m^2/s$ and $\tau=6 \times 10^4 s$ are taken from the fig(1F) in \cite{Wu14PNAS}) is high compared to brownian diffusion $D_{Br}=k_B T/\gamma=6.67 \times 10^{-20} m^2/s$ in the matrix. The Brownian diffusion in water is $5 \times 10^{-14} m^2/s$. The viscosity of gel is greater than the water. Therefore the diffusion due to self motility of cell is always greater than the Brownian diffusion. The scaling is determined by the equation (4). In the numerator, $D_0=2(Dk^2\mu(k)\phi_0+(f_0^2 k^2 \xi(\omega) )\phi_0)$ term is dominated by self-motility term $(f_0^2 k^2 \xi(\omega) )\phi_0)$ by following the above argument. The resulting superdiffusion exponents are reported in the main text. The non-linear contributions are included in the self-energy term $\Sigma$.
The non-linear interactions determine the dynamical behavior. The time-dependent adhesion interactions and long-range elastic interactions determine the superdiffusive behavior for the mesenchymal phenotype. On the other hand, long-range hydrodynamic interactions determine superdiffusive behavior. We can tune the strength of the interactions that will change the magnitude of the coefficient of the non-linear term. The scaling of the non-linear term remains unchanged. Therefore the degree of superdiffusion is unchanged.}

\section{Scaling exponents}
In a self consistent mode coupling theory, we now replace $\nu$ by $\Delta \nu$ in 
the self energy term $\Sigma({0},\omega, \omega_{\tau_f})$ in the first term in Fig.(\ref{fig:rg6}), use $G\sim \omega_{\tau_f}^{-1}$  as from 
Eq.~(\ref{green}) and $C$, which follows from the FDT. 
{According to scale transformation, we know that $\omega \sim k^z$, $\omega_\tau \sim k^{4z-2}$, $G \sim k^{-4z+2}$, $C \sim k^{-8z+4}$ and the vertex factor $V \sim k^{4z-2}$. The self energy term in Fig.(\ref{fig:rg6}) can be written as 
$\Sigma({0},\omega, \omega_{\tau_f})\sim  \int \frac{d^d {\bf k'}}{(2\pi)^d} \frac{d\omega'}{2\pi} \frac{d\omega'_\tau}{2\pi} V V GC$.}
By carrying out the momentum count of $\Sigma({0},\omega, \omega_{\tau_f})$, and 
using $\Delta \nu \sim k^z$, we find that $\Sigma({\bf k},\omega, \omega_{\tau_f})\sim k^{d-z}$. 
Using Eq. (\ref{scale}) and $\nu/D_0 \sim k^{3z-2}$, we have $k^{4z-2}\sim k^{d+z}$, which leads to $z=\frac{d+2}{3}$. Where, $\nu \sim a(\omega)\approx \omega =k^z$ and use $\lambda^{-1}$ is small because in the case of flow, cells are amoeboid phenotype with short lived adhesion with collagen fiber. 
MSD exponent $\alpha=2/z=6/(d+2)$. In 3D, $\alpha=1.2$, i.e., the amoeboid cells undergo super-diffusion.

\section{The expression for $\Sigma({\bf l},\omega, \omega_{\tau_f})$}
\begin{eqnarray}\label{selfenergy}
&&\Sigma({\bf k},\omega, \omega_{\tau_f})=\frac{2}{(Dk^2\mu(k)\phi_0+(f_0^2 k^2 \xi(\omega) )\phi_0))} \\ \nonumber && \int _{\bf k',\omega',\omega'_\tau}   V({\bf k},\omega,{\bf k'},\omega') V({\bf k},\omega,{\bf k}-{\bf k'},\omega-\omega')
\\ \nonumber &&G({\bf k'},\omega', \omega'_\tau) C({\bf k-k'},\omega-\omega',\omega_{\tau_f}-\omega'_\tau),
\end{eqnarray}

where veterx term,
 $V({\bf k},\omega,{\bf k'},\omega')=\{i \omega+Dk^2 \mu(k)+\phi_0 k^2 \mu(k)a(\omega)U({\bf k})\} 
  \{(-{\bf k'} \cdot {\bf k})\mu({\bf k'})a(\omega) U({\bf k'})\}+ 
 \{i \omega'+Dk'^2 \mu(k')+\phi_0 k'^2 \mu(k')U({\bf k'})+k'^2 \mu(k')\} 
 \{(-{\bf k'} \cdot {\bf k})\mu({\bf k})a(\omega) U({\bf k})\} +
 \{i \omega'+Dk'^2 \mu(k')+\phi_0 k'^2 \mu(k')a(\omega)U({\bf k'})\} 
 \{(-{\bf k'} \cdot ({\bf k}-{\bf k'}))\mu({\bf k}-{\bf k'})a(\omega) U({\bf k}-{\bf k'})\}$

\end{document}